\documentclass[5p,authoryear,times,twocolumn]{elsarticle}
\usepackage{natbib}
\usepackage{color}
\usepackage[latin1]{inputenc}
\usepackage{graphics}
\usepackage{booktabs}
\usepackage{amssymb}
\usepackage{overpic}
\journal{Icarus}
\begin{document}
\begin{frontmatter}
\title{Micrometer-sized ice particles for planetary-science experiments - I. Preparation, critical rolling friction force, and specific surface energy}
\author[igep]{B. Gundlach}
\ead{b.gundlach@tu-bs.de}
\author[igep]{S. Kilias}
\author[igep]{E. Beitz}
\author[igep]{J. Blum}
\address[igep]{Institut für Geophysik und extraterrestrische Physik, Technische Universität Braunschweig, \\Mendelssohnstr. 3, D-38106 Braunschweig, Germany}
\begin{abstract}
Coagulation models assume a higher sticking threshold for micrometer-sized ice particles than for micrometer-sized silicate particles. However, in contrast to silicates, laboratory investigations of the collision properties of micrometer-sized ice particles (in particular, of the most abundant water ice) have not been conducted yet. Thus, we used two different experimental methods to produce micrometer-sized water ice particles, i. e. by spraying water droplets into liquid nitrogen and by spraying water droplets into a cold nitrogen atmosphere. The mean particle radii of the ice particles produced with these experimental methods are $(1.49 \pm 0.79) \, \mathrm{\mu m}$ and $(1.45 \pm 0.65) \, \mathrm{\mu m}$. Ice aggregates composed of the micrometer-sized ice particles are highly porous (volume filling factor: $\phi = 0.11 \pm 0.01$) or rather compact (volume filling factor: $\phi = 0.72 \pm 0.04$), depending on the method of production. Furthermore, the critical rolling friction force of $F_{Roll,ice}=(114.8 \pm 23.8) \times 10^{-10}\, \mathrm{N}$ was measured for micrometer-sized ice particles, which exceeds the critical rolling friction force of micrometer-sized $\mathrm{SiO_2}$ particles ($F_{Roll,SiO_2}=(12.1 \pm 3.6) \times 10^{-10}\, \mathrm{N}$). This result implies that the adhesive bonding between micrometer-sized ice particles is stronger than the bonding strength between $\mathrm{SiO_2}$ particles. An estimation of the specific surface energy of micrometer-sized ice particles, derived from the measured critical rolling friction forces and the surface energy of micrometer-sized $\mathrm{SiO_2}$ particles, results in $\gamma_{ice} = 0.190 \, \mathrm{J \, m^{-2}}$.
\end{abstract}
\end{frontmatter}

\section{Introduction}
Coagulation of micrometer-sized particles plays an important role in molecular clouds \citep{Ossenkopf1993, Weidenschilling1994, Ormel2009} and protoplanetary disks \citep{BlumWurm2008, Zsom2010}. In both environments, the dominating dust materials are silicates, carbonaceous material, and ices. A wide variety of experimental investigations of the coagulation and fragmentation of aggregates composed of micrometer-sized dust particles, mostly silicates, have been performed \citep{BlumWurm2008,Guettler2010}. Micrometer-sized silicate grains stick to one another for collision velocities $\lesssim 1\,\mathrm{m\,s^{-1}}$ \citep{Guettler2010}. Dust collisions can lead to fragmentation of the dust aggregate if the collision velocity exceeds the fragmentation threshold for aggregates, which is expected for velocities in the range of $\sim(1-100) \, \mathrm{m\, s^{-1}}$, depending on grain size and material. This effect has been investigated in the laboratory for silicates but not for ices.
\par
Water ice obviously played an important role in planet formation of our own Solar System as can be seen by the high $\mathrm{H_2O}$-ice abundances in the planetary bodies of the outer Solar System. It is generally assumed that the increased stickiness of water-ice particles over the refractory materials plays a crucial role in the rapid formation of the giant planets and their satellite systems. Unfortunately, most of the planetary bodies of the outer Solar System lost memory of their formation processes, due to sintering, melting, or high-pressure effects. An exception are probably the comets, which never experienced enhanced pressures or temperatures since their formation in the outer reaches of the solar nebula (but see also \citet{Levison2010} for an extrasolar formation of part of the comets). Here, we expect the building blocks of the planetary bodies to be preserved in their original state. The first direct evidence for micrometer-sized ice particles on a cometary nucleus was given by the Deep Impact mission, when the ejected particles of the comet 9P/Tempel were observed in the infrared \citep{Sunshine2007}. Recently,  A'Hearn and coleagues (pers. comm.) found evidence for larger icy bodies (presumably aggregates of microscopic ice grains) during the flyby of comet Hartley 2. Thus, we believe that understanding the collision and adhesion behavior of ice particles and the subsequent formation of ice aggregates can help to better understand the physical processes which led to the formation of the ice-dominated bodies of the outer Solar System.
\par
In contrast to silicates, the collision properties of ices (and, in particular, of the most abundant water ice) are largely undetermined. Coagulation models assume a higher surface energy and therefore a higher sticking threshold for ice particles \citep{Wada2007,Wada2008} as compared to silicates. However, experimental investigations of the specific surface energy of water ice are scarce. Furthermore, experiments on the coagulation or fragmentation behavior of microscopic ice particles under astrophysical conditions have not been conducted yet.
\par
Laboratory experiments with macroscopic ice particles in the cm- to dm-regime were performed by \citet{Hatzes1988}, \citet{Hatzes1991}, \citet{Supulver1997}, and \citet{Heisselmann2010}. In all these experiments, no sticking of the ice particles was observed, even at velocities below $0.1\,\mathrm{mm s^{-1}}$. The existence of a frost layer increased the sticking threshold of the macroscopic ice particles to observable values. However, it is questionable whether these experiments can be used to derive the collision properties of microscopically small ice particles.
\par
In this paper, we will present two experimental methods to produce micrometer-sized ice particles and ice aggregates (see Sect. \ref{Preparation of micrometer-sized ice particles}). In Sect. \ref{Characterization of micrometer-sized ice particles}, the size distribution of the produced micrometer-sized ice particles, the volume filling factor of the ice aggregates, the critical rolling friction force and the specific surface energy of the micrometer-sized ice particles are investigated. Finally, a summary of the obtained results and an outlook on future experiments are given in Sect. \ref{Conclusion and outlock}.

\section{Preparation of micrometer-sized ice particles}\label{Preparation of micrometer-sized ice particles}
\begin{figure}[t!]
\centering
\includegraphics[angle=0,width=1.0\columnwidth]{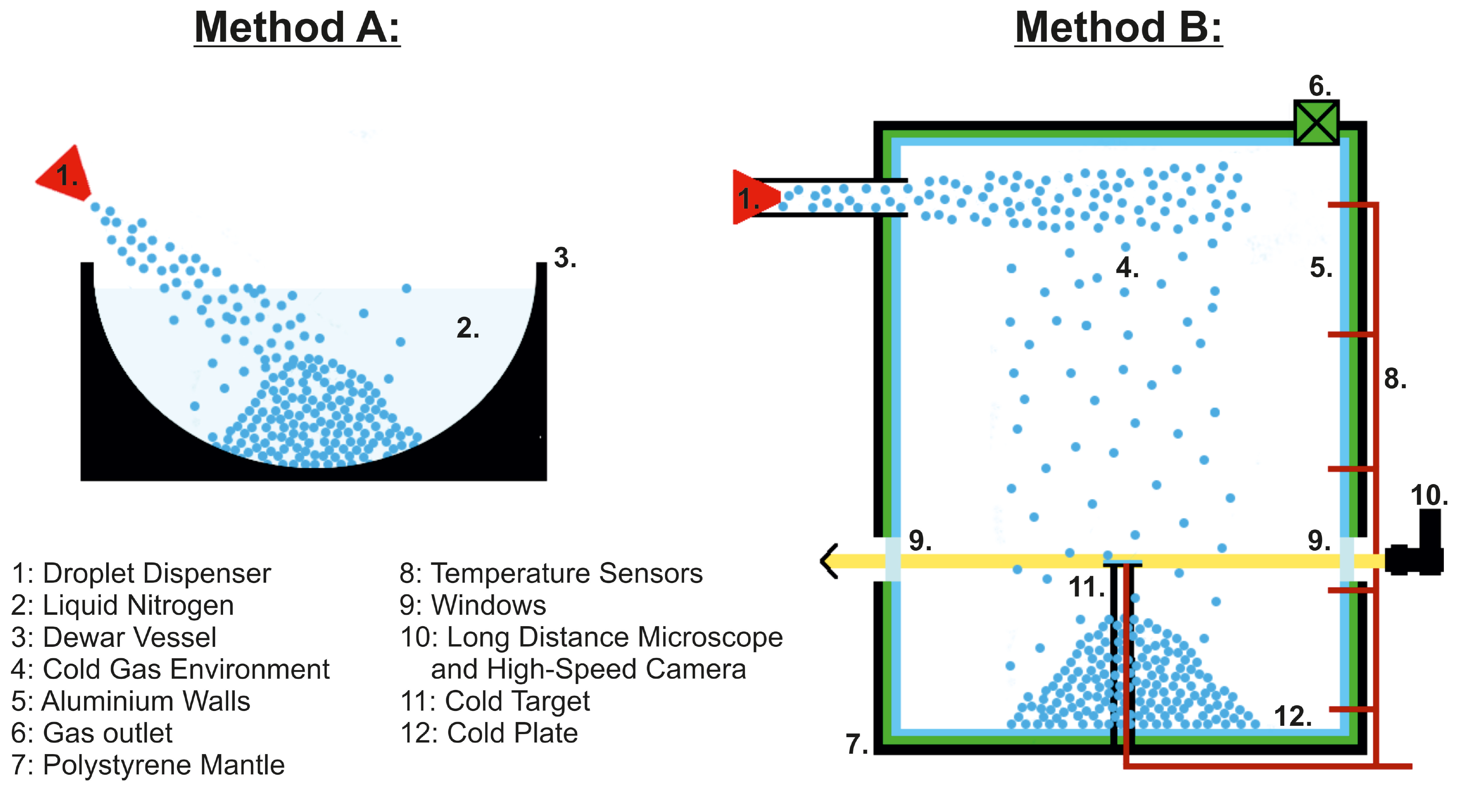}
\caption{Design of the two different experimental setups (A and B), which were established to produce micrometer-sized ice particles.}
\label{ExpMethods}
\end{figure}
\begin{figure}[t]
\centering
\includegraphics[angle=90,width=0.9\columnwidth]{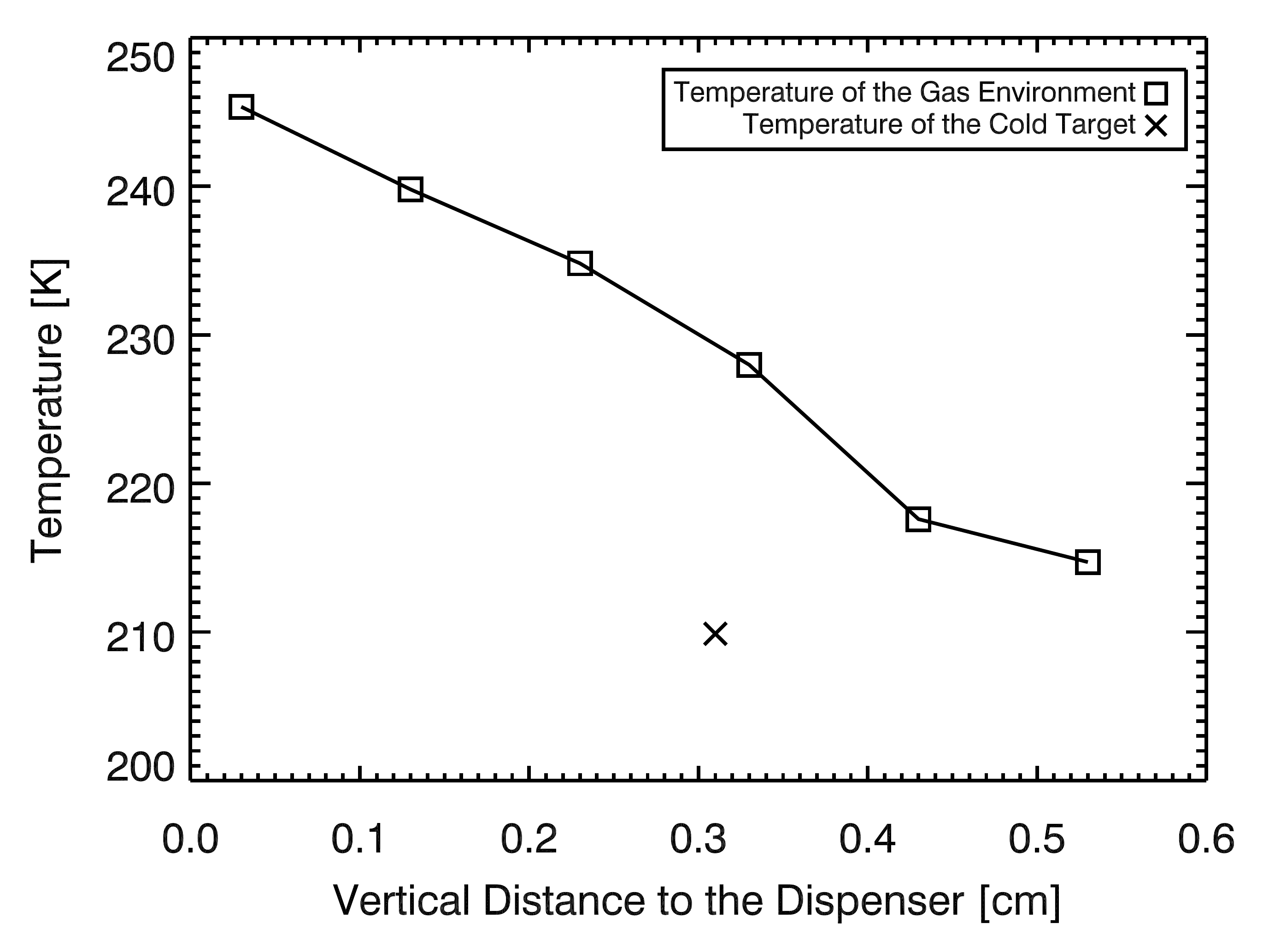}
\caption{Temperature distribution of the dry gas environment (squares) and the cold target (cross) inside the experimental chamber of setup B. The solid lines are included to guide the eye.}
\label{TempDistr}
\end{figure}
Two different experimental setups (A and B, see Fig. \ref{ExpMethods}) were developed to produce ice particles with diameters ranging from sub-micrometer-size to several micrometers (see Sect. \ref{Size distribution}). In both experiments, liquid water was dispersed by a commercial droplet dispenser (1, in Fig. \ref{ExpMethods}). For instant cooling to $\sim 77 \, \mathrm{K}$ (setup A), the dispersed water was directly sprayed into liquid nitrogen (2), which was stored in a dewar vessel (3). Afterwards, separation of the produced ice particles from the liquid nitrogen was conducted by filtration or by evaporation of the liquid nitrogen.
\par
The second experiment (setup B) was performed by spraying the dispersed water droplets into a dry, cold gas environment (4). Liquid nitrogen was filled into the experimental setup before the production of the ice particles was started in order to cool the aluminum walls (5) of the experimental setup. The dry cold gas environment was generated by the evaporated nitrogen, which was kept cool by the aluminum walls after the nitrogen had fully evaporated. A gas outlet (6) was incorporated into the experimental setup to enable the escape of the nitrogen vapor. Due to the strong upward directed gas flux of the evaporating nitrogen, the production of the ice particles was started after the liquid nitrogen had evaporated. For safety reasons, the cold experiment was thermally isolated by a polystyrene mantle (7). Temperature sensors (8) were positioned inside the experiment to monitor the vertical temperatures of the gas environment. A typical temperature distribution inside the experimental chamber during an experiment is shown in Fig. \ref{TempDistr}. The measured temperature of the cold target is lower than the temperature of the gas environment, due to (a) the connection of the cold target with the cold aluminum walls of the experimental chamber and (b) the heat deposition of the warm water droplets to the gas.
\par
Formation of ice particles occurred during sedimentation of the water droplets inside the dry, cold gas environment. The falling speed of micrometer-sized particles at atmospheric pressure is about  $300 \, \mathrm{\mu m \, s^{-1}}$. Thus, the duration of sedimentation inside the dry, cold gas environment is much longer than the required time to freeze the dispersed water. Two windows (9) were incorporated into the aluminum wall, $31\,\mathrm{cm}$ beneath the droplet dispenser, to illuminate and observe the sedimenting ice particles with a long-distance microscope together with a high-speed camera (10). For the measurement of the critical rolling friction force of micrometer-sized ice particles (see Sect. \ref{Rolling friction}), a cold target (11) was positioned in the field of view of the long-distance microscope. After sedimentation, the ice particles were stored on a cold plate (12) at the bottom of the chamber, $53 \, \mathrm{cm}$ beneath the droplet dispenser, for further investigations.

\section{Characterization of the produced micrometer-sized ice particles}\label{Characterization of micrometer-sized ice particles}
\subsection{Size distribution}\label{Size distribution}
\begin{figure}[t]
\centering
\includegraphics[angle=0,width=1.0\columnwidth]{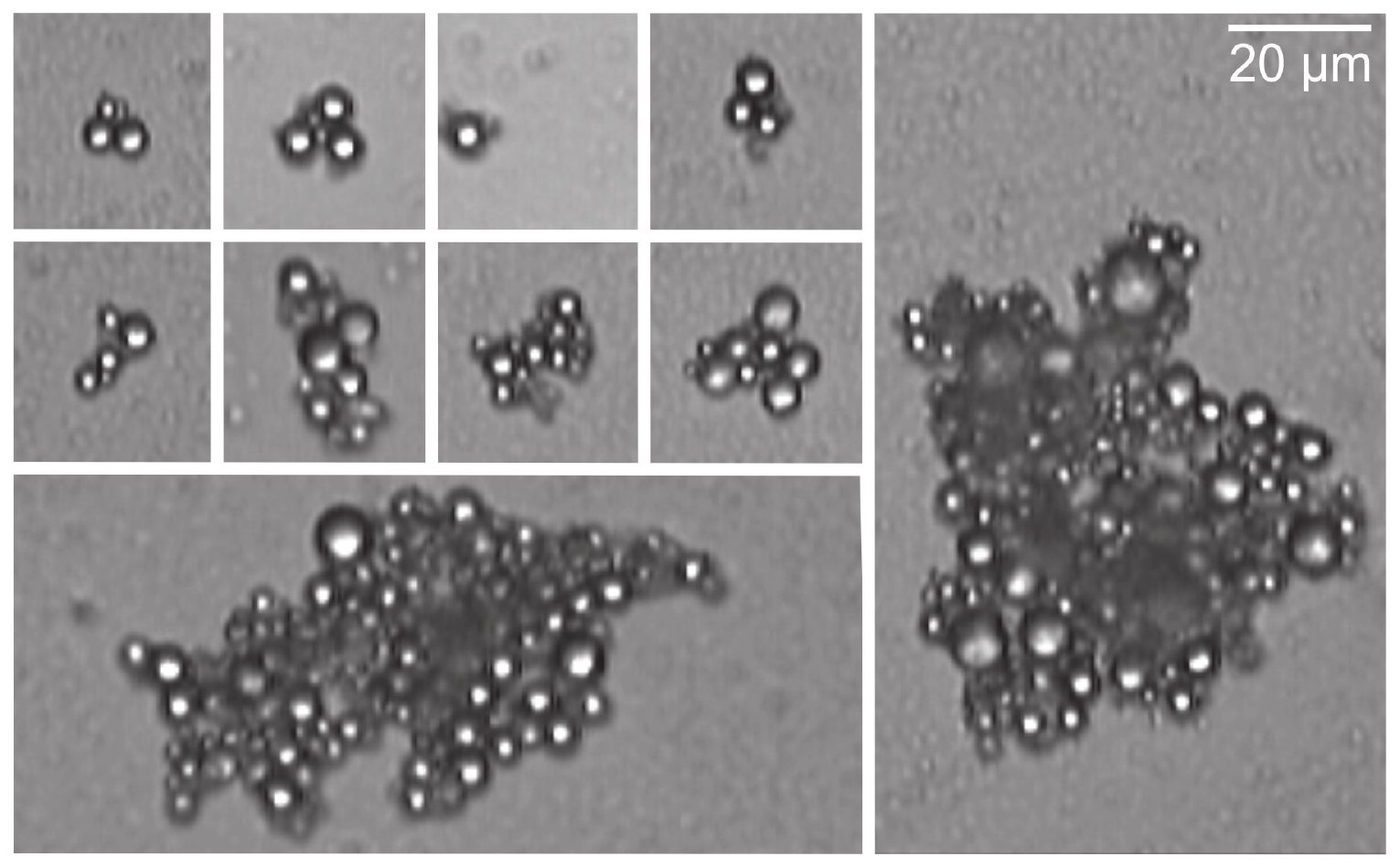}
\caption{Ice particles produced with setup A and observed with the light microscope.}
\label{fig_particles}
\end{figure}

The size distributions of the produced ice particles were investigated for both experimental methods. For the size estimation of the ice particles produced with setup A, a light microscope was used. The ice particles were positioned on a cooled object slide. To avoid condensation of moisture (frost) on the ice particles as well as on the optical components, the light microscope was positioned inside a glove box which was filled with dry nitrogen gas. The prevention of water vapor condensation is very important, because the formation of frost on ice particles can have a strong influence on their sticking properties, like e.g. on the coefficient of restitution (see e. g. \citet{Heisselmann2010, Hatzes1988}). To avoid melting of the ice particles during the measurements, the sample holder of the microscope was cooled and the illumination of the microscope was modified by an infrared filter to block the infrared radiation, due to the strong light absorption of water ice in the infrared. Fig. \ref{fig_particles} shows ice particles produced with setup A and observed with the light microscope.
\par
\begin{figure}[h]
\centering
\includegraphics[angle=0,width=0.9\columnwidth]{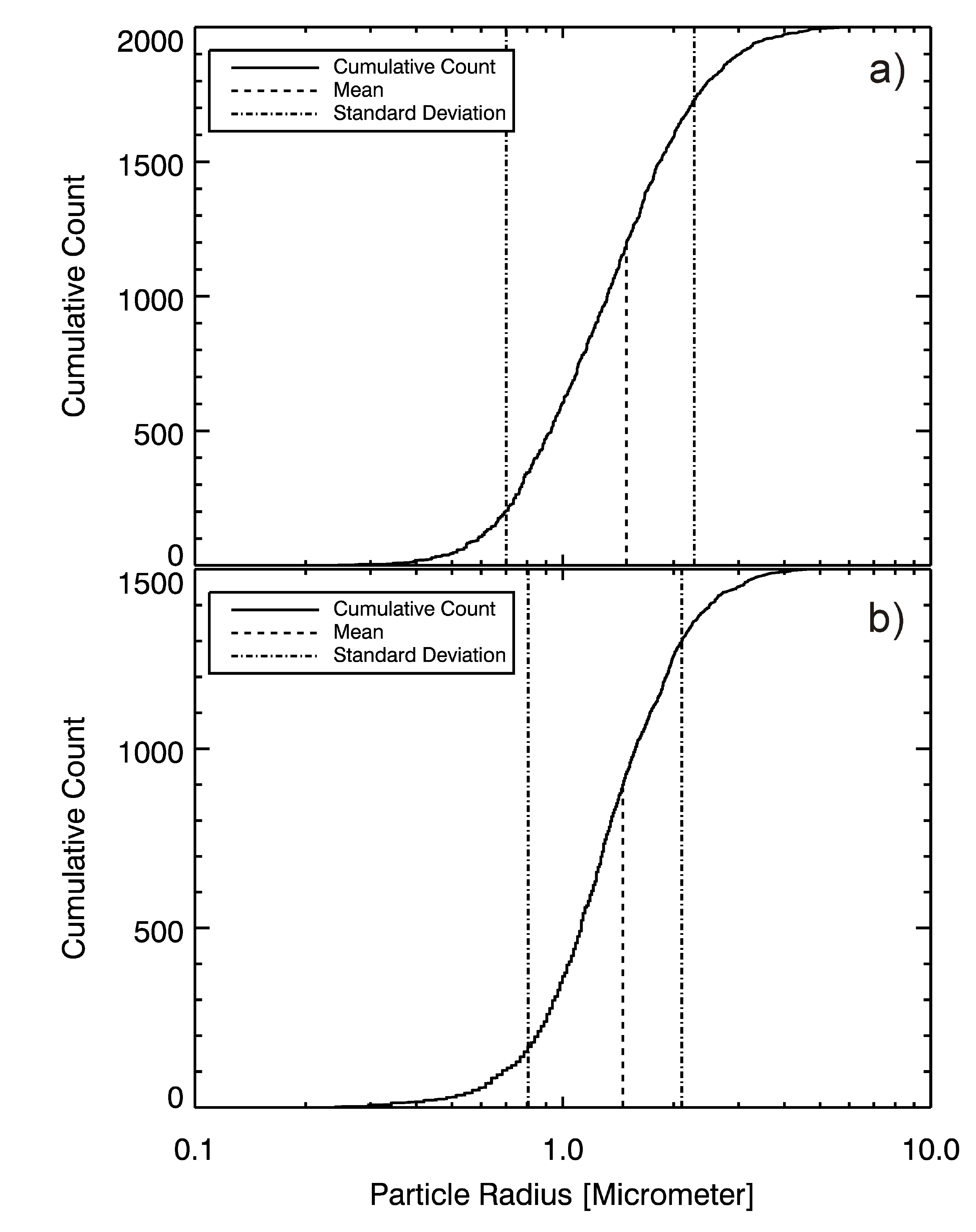}
\caption{Cumulative size distributions of the ice particles produced with setup A (a, solid curve) and produced with setup B measured at the position of the cold target (b, solid curve). The arithmetic mean radii of the particles (dashed lines), which were produced using the different setups, are very similar: $a_{0A} = (1.49 \pm 0.79) \, \mathrm{\mu m}$ (setup A) and $a_{0B} = (1.45 \pm 0.65) \, \mathrm{\mu m}$ (setup B). The errors of both measurements are given by the standard deviations of the measurements (dashed dotted lines).}
\label{fig_particles_size}
\end{figure}
The measured cumulative size distribution of the ice particles is visualized in Fig. \ref{fig_particles_size}a (solid curve). Ice particles produced with setup A have an arithmetic mean radius of $a_{0A} = (1.49 \pm 0.79) \, \mathrm{\mu m}$ (dashed line), the error indicates the standard deviation of the measured radii (dashed dotted line). The particle radii range from $a_{A,min} = 0.24\,\mathrm{\mu m}$ to $a_{A,max} = 6.07\,\mathrm{\mu m}$.
\par
An estimation of the size distribution of the ice particles produced with setup B was carried out using the long-distance microscope (see Sect. \ref{Preparation of micrometer-sized ice particles}). The measurement of the size distribution was performed at the height of the cold target. Fig. \ref{fig_particles_size}b shows the cumulative size distribution of the ice particles produced with setup B (solid curve). The particle radii range from $a_{B,min} = 0.24\,\mathrm{\mu m}$ to $a_{B,max} = 5.52\,\mathrm{\mu m}$. Furthermore, the arithmetic mean radius of the ice particles produced with setup B is $a_{0B} = (1.45\pm0.65)\, \mathrm{\mu m}$ (dashed line), with the error of the measurement given by the standard deviation of the measurement of radii (dashed dotted line). Both experimental setups produced very similar micrometer-sized ice particles, due to the usage of the same commercial droplet dispenser. Examples of sedimenting ice particles, observed with the long-distance microscope, are shown in Fig. \ref{AggCol}.

\subsection{Volume filling factor of ice aggregates composed of micrometer-sized ice particles}\label{Porosity of ice aggregates composed of micrometer-sized ice particles}
\begin{figure}[b!]
\centering
\includegraphics[angle=0,width=1.0\columnwidth]{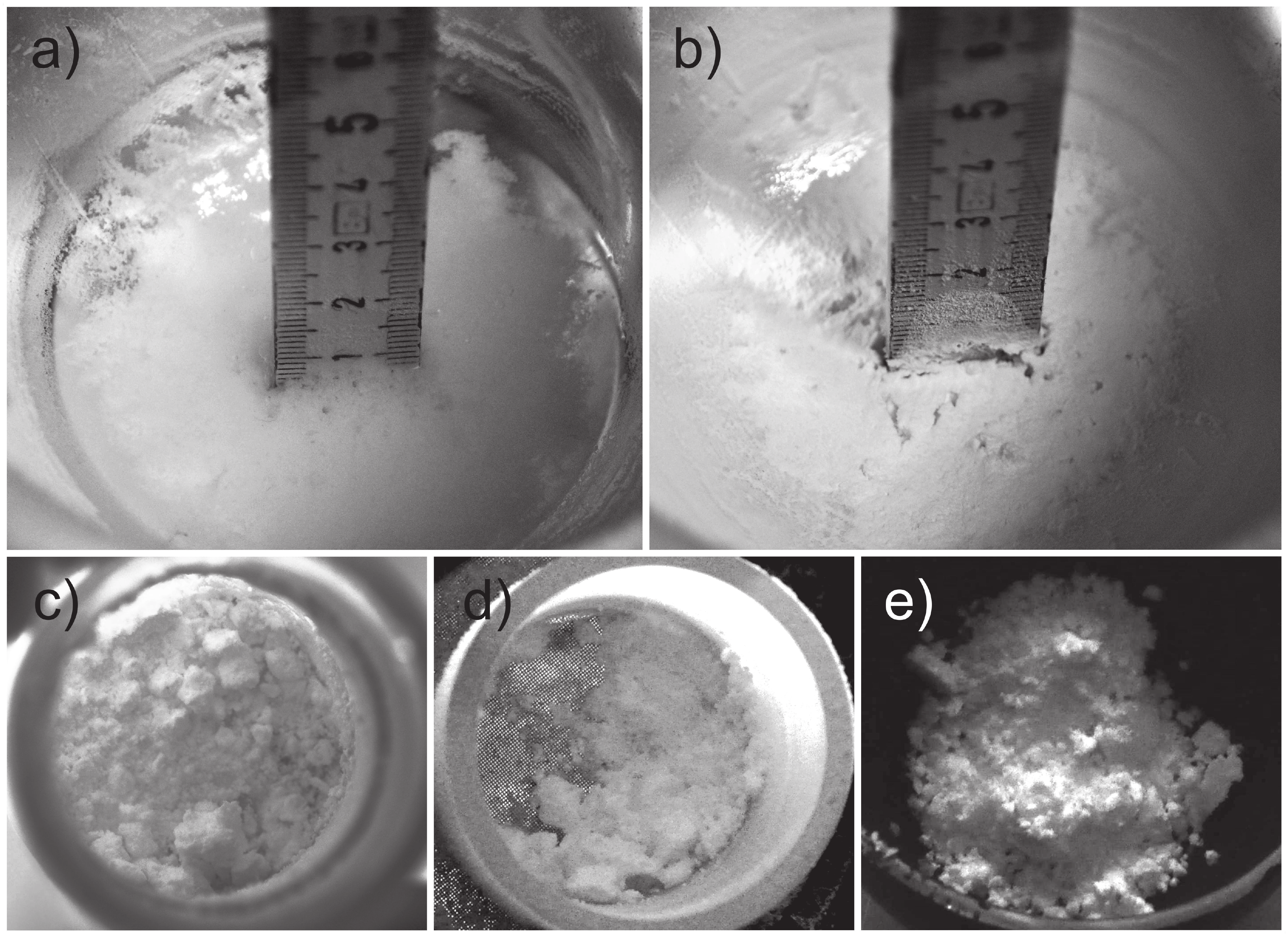}
\caption{Images of ice aggregates composed of micrometer-sized ice particles. Ice aggregates, produced with setup A, were separated from the liquid nitrogen by evaporation of the liquid nitrogen or by filtration. The images are showing ice aggregates before and after evaporation of the liquid nitrogen (a and b) and ice aggregates after filtration (c and d). An ice aggregate produced with setup B is shown in image (e).}
\label{Eisaggregate}
\end{figure}

Ice aggregates composed of micrometer-sized ice particles were produced with both experimental setups (A and B). Fig. \ref{Eisaggregate} shows examples of different ice aggregates produced by the two setups. Measurements of the volume filling factor of these aggregates were conducted, determining the mass as well as the occupied volume of the aggregates before and after melting.
\begin{figure*}[t!]
\centering
\includegraphics[angle=0,width=1.0\textwidth]{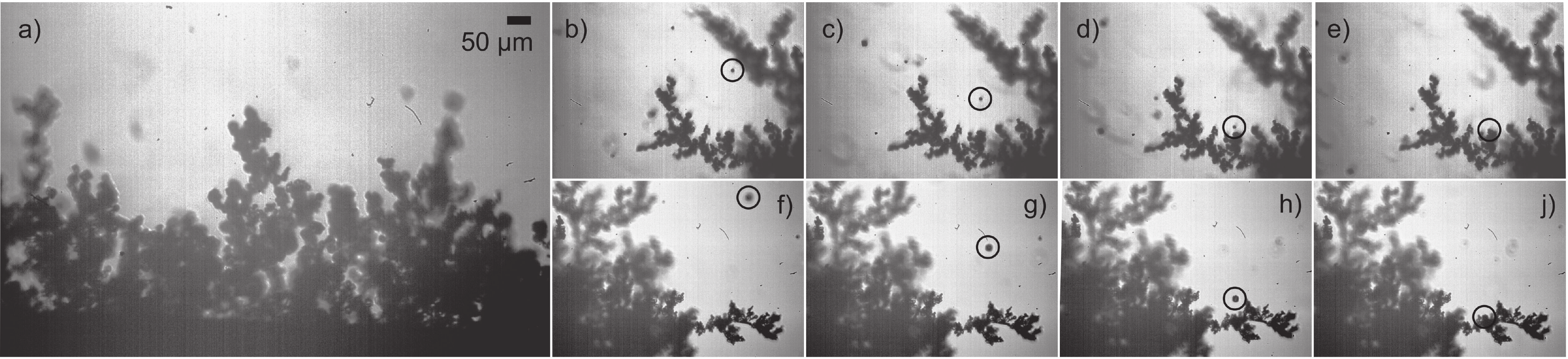}
\caption{Sedimenting ice particles and ice aggregates during formation (setup B), observed with the long-distance microscope. The general structure of these very porous aggregates, $\phi_B = 0.11 \pm 0.01$, is shown in a. This high porosity is caused by the "hit-and-stick" behavior of micrometer-sized particles at low impact velocities. Examples of two "hit-and-stick" collisions between micrometer-sized ice particles are shown in the image sequences b-e and f-j.}
\label{AggCol}
\end{figure*}
\par
A volume filling factor of $\phi_A = (0.72 \pm 0.04)$ was measured for five ice aggregates, which were produced by spraying the dispersed water into liquid nitrogen (setup A). The error is given by the statistical variation of these measurements. The determined value is close to the volume filling factor of hexagonal close packed material: $\phi_{hcp} = \sim 0.74$. Thus, ice aggregates produced with setup A are relatively compact, due to the sedimentation and rearrangement of the ice particles in the liquid nitrogen.
\par
However, ice aggregates produced with setup B, are highly porous, $\phi_B = 0.11 \pm 0.01$. To calculate the mean volume filling factor of ice aggregates produced with setup B, we deposited the ice particles on a cooled metal plate (with dimensions $45\, \mathrm{mm}$ $\times$ $45\, \mathrm{mm}$) inside the experimental chamber. The occupied volumes of 15 ice aggregates were determined by measuring the heights of the deposited ice aggregates with the long-distance microscope (see Sect. \ref{Preparation of micrometer-sized ice particles}) and taking the  cross-section area of the metal plate into account. Afterwards, the ice aggregates were rapidly weighted to avoid condensation of frost and the volume filling factors of the ice aggregates were calculated by comparing the mass of an individual ice aggregate with the mass of solid water ice occupying the same volume as the ice aggregate at the same temperature.
\par
An example of the grown ice aggregates, observed with the long-distance microscope, is shown in Fig. \ref{AggCol}a. This image demonstrates the high porosity of the ice aggregates, which can be explained with "hit-and-stick" behavior of the sedimenting micrometer-sized particles at low impact velocities. Two different "hit-and-stick" collisions of micrometer-sized ice particles with the grown ice aggregate are shown in the image sequences \ref{AggCol}b-e and \ref{AggCol}f-j. Sticking at the first point of contact of micrometer-sized particles was also found in previous works with similar setups \citep{BlumWurm2000}, in which the sticking properties of micrometer-sized $\mathrm{SiO_2}$ particles, with a radius of $a_0 = 0.95\, \mathrm{\mu m}$, were investigated.

\subsection{Critical rolling friction force and specific surface energy}\label{Rolling friction}
The critical rolling friction force between two adhering particles is an important quantity for the characterization of their collisional properties and of the restructuring of particle aggregates in contact \citep{DominikTielens1995, Heim1999, BlumWurm2000, Wada2008}. An expression for the critical rolling friction force between two spheres was calculated by \citet{DominikTielens1995},
\begin{equation}
F_{Roll} \, = \, 6 \, \pi \gamma \, \xi \, \mathrm{,}
\label{eq0}
\end{equation}
where $\gamma$ is the specific surface energy of the material and $\xi$ is the critical rolling displacement of the particle, which is the distance one sphere may roll over the other before irreversible rearrangement in the contact zone occurs. The critical rolling friction force of uncoated monodisperse silica particles of $a_0 = 0.95 \, \mathrm{\mu m}$ radius was experimentally investigated by \citet{Heim1999}, by pulling on chainlike aggregates with an AFM tip and testing their resistance to a forced oscillating motion on one end of the chain, while the other end was fixed. From this experiment, a mean critical rolling friction force of $F_{Roll,SiO_2} = (8.5 \pm 1.6 ) \times 10^{-10}\, \mathrm{N}$ was found for uncoated $\mathrm{SiO_2}$ particles.
\par
Another experimental approach for the estimation of the critical rolling friction force of monodisperse $\mathrm{SiO_2}$ particles with the same size was performed by \citet{BlumWurm2000}. In this work, the $\mathrm{SiO_2}$ particles were coated with a silicon-organic mantle (dimethyldimethoxysilane, $\mathrm{(CH_3)_2Si(OCH_3)_2}$), to guarantee a nonpolar, hydrophobic surface layer. They found that the additional mantle enhanced the specific surface energy of the $\mathrm{SiO_2}$ particles by a factor of $1.35$. Thus, the obtained critical rolling friction force of the coated $\mathrm{SiO_2}$ particles was corrected by this factor for a comparison with uncoated $\mathrm{SiO_2}$ particles. The critical rolling friction force was estimated by observing several collisons between $\mathrm{SiO_2}$ particles and agglomerates composed of individual $\mathrm{SiO_2}$ particles, in which gravitational restructuring was manifested through a slow morphological transition within the aggregate layer. In this case, the critical
rolling friction force can be calculated as follows,
\begin{equation}
F_{Roll} \, = \, m \, g_0 \, \frac{a_{cm}}{a_0} \, \mathrm{,}
\label{eq1}
\end{equation}
where $m$ is the mass of the restructuring aggregate, $g_0 = 9.81 \, \mathrm{m \, s^{-2}}$ is the gravitational acceleration of the Earth, $a_0$ is the mean particle radius and $a_{cm}$ is the horizontal projection of the distance from the center of gravity of the rotating aggregate to the point of contact about which restructuring occurs. The analysis of several temporally resolved restructuring events yielded a mean critical rolling friction force of $F_{Roll,SiO_2}  = (5.0 \pm 2.5) \times 10^{-10}\, \mathrm{N}$ for the coated $\mathrm{SiO_2}$ particles, which was corrected to $F_{Roll,SiO_2}  = (3.7 \pm 1.9) \times 10^{-10}\, \mathrm{N}$ for a comparison with the results for uncoated $\mathrm{SiO_2}$ particles.
\par
\begin{figure*}[t]
\centering
\includegraphics[angle=0,width=1.0\textwidth]{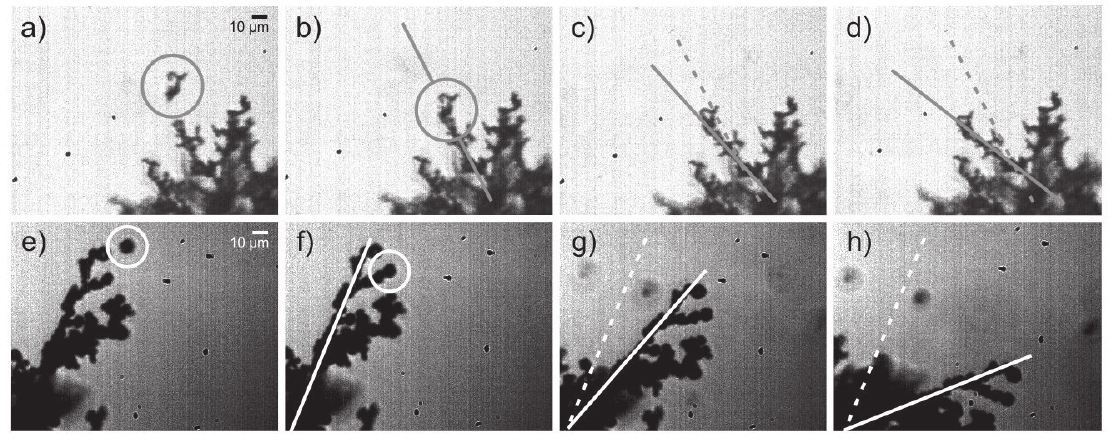}
\caption{Examples of time resolved restructuring events of aggregates composed of micrometer-sized $\mathrm{SiO_2}$ particles (a-d) or $\mathrm{H_2O}$ ice particles (e-h). The restructuring events were initiated by the addition of an impacting particle or cluster. The images were taken every $0.03\,\mathrm{s}$ (a-d) and every $0.02\,\mathrm{s}$ (e-h).}
\label{AbknAst}
\end{figure*}
To test the capability of our experimental setup, we measured the critical rolling friction force of uncoated, monodisperse $\mathrm{SiO_2}$ particles of $a_0 = 0.75 \, \mathrm{\mu m}$, using the experimental setup B (see Sect. \ref{Preparation of micrometer-sized ice particles}) together with the procedure introduced by \citet{BlumWurm2000}. In this case, the dust was dispersed by a commercial dust-disperser and sprayed into the experimental chamber. An analysis of the size distribution of the dispersed dust showed that the dust is mostly dispersed into single grains ($\sim75 \, \%$) and into cluster of two particles ($\sim15\, \%$) as well as of three particles ($\sim 10 \,\%$). Our measurements yielded a critical rolling friction force of $F_{Roll,SiO_2}  = (12.1 \pm 3.6) \times 10^{-10}\, \mathrm{N}$, which is relatively close (within one standard deviation) to the quantity measured by \citet{Heim1999}, but slightly further away from the corrected value obtained by \citet{BlumWurm2000}. Fig \ref{AbknAst}a-d show an example of a time-resolved restructuring event of an aggregate composed of micrometer-sized $\mathrm{SiO_2}$ particles. A comparison of the different measured quantities is given in Table \ref{TabComp}. However, a better reproduction of the previously published values is not possible, due to the uncertainty of the mass estimation of the restructuring particle chains, which have occurred in this work as well as in the previous work conducted by \citet{BlumWurm2000}. Nevertheless, the obtained critical rolling friction force is in the expected range between $5 \times 10^{-11} \, \mathrm{N}$ and $4 \times 10^{-9} \, \mathrm{N}$ \citep{Heim1999}.
\begin{figure}[t]
\centering
\includegraphics[angle=180,width=0.9\columnwidth]{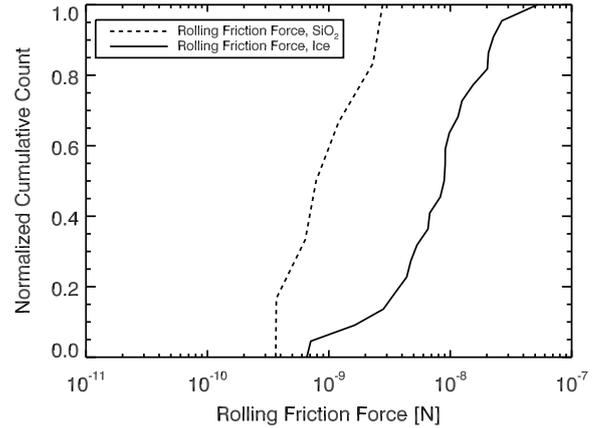}
\caption{Normalized cumulative count of the critical rolling friction force measured for micrometer-sized $\mathrm{SiO_2}$ particles (dashed curve) as well as for micrometer-sized ice particles (solid curve). Mean critical rolling friction forces of $F_{Roll,SiO_2}  = (12.1 \pm 3.6) \times 10^{-10}\, \mathrm{N}$ and of $F_{Roll,ice} = (114.8 \pm 23.8) \times 10^{-10}\, \mathrm{N}$ were measured for the micrometer-sized $\mathrm{SiO_2}$ and ice particles, respectively.}
\label{CumFroll}
\end{figure}
\par
\begin{table*}[t!]
\footnotesize
\begin{center}
    \caption{Comparison of the critical rolling friction force of micrometer-sized $\mathrm{SiO_2}$ and ice particles measured by \citet{Heim1999}, \citet{BlumWurm2000} and in this work. The errors of the critical rolling friction force are given by the statistical uncertainties of the measurements.}\vspace{2mm}
    \begin{tabular}{lcccc}
        \toprule
        Reference                                                   & Material        & $a_0$ [$\mathrm{\mu m}$]          &     $\mathrm{F_{Roll}}$ [$\times 10^{-10}$ N]  &     $\gamma$ [$\mathrm{J\,m^{-2}}$] \\
        \midrule
        \citet{Heim1999}                                                & uncoated $\mathrm{SiO_2}$  & $0.95$          & $8.5 \pm 1.6$ &  0.014   \\[0.6mm]
        \citet{BlumWurm2000}$^\dag$                                             & coated $\mathrm{SiO_2}$  & $0.95$          & $3.7 \pm 1.9$ &   0.019   \\[0.6mm]
         This work                                                  & uncoated $\mathrm{SiO_2}$  & $0.75$          & $12.1\pm3.6$  &     0.020       \\[0.6mm]
         This work                                                  & $\mathrm{H_2O \ ice}$    & $1.45\pm0.65$   & $114.8 \pm 23.8$  &    0.190    \\
        \bottomrule
        \multicolumn{4}{l}{$\dag$: Corrected critical rolling friction force, due to the additional coating of the $\mathrm{SiO_2}$ particles.}
    \end{tabular}
     \label{TabComp}
     \end{center}
\end{table*}
For comparison, the critical rolling friction force of the produced micrometer-sized ice particles was also investigated with the same method. In total, 23 temporally resolved restructuring events (see Fig. \ref{AbknAst}e-h) of ice particles on the cold target were analyzed. The critical rolling friction force of the ice particles was then calculated, taking a mean radius of $a_{0B} = 1.45\, \mathrm{\mu m}$, a mean cross section of a single ice sphere of $ S = (8.90 \pm 0.24) \times 10^{-12} \,\mathrm{m^2}$ and a mean mass of $m = (2.65\pm 0.13) \times 10^{-14}\,\mathrm{kg}$ of the ice particles into account. These values were derived from the cumulative size, cross section and mass distributions of the ice particles. A critical rolling friction force of $F_{Roll,ice} =(114.8 \pm 23.8) \times 10^{-10}\, \mathrm{N}$ was derived for the ice particles, in which the uncertainty of the measurement is given by the statistical error of the mean value. The temperature of the cold target during the measurements was in the range between $189 \, \mathrm{K}$ and $226 \, \mathrm{K}$, which is relatively high, compared to the temperature in molecular clouds or protoplanetary discs. Therefore, future experiments should be carried out at lower temperatures in order to estimate the temperature dependence of the attractive bonding between micrometer-sized ice particles.
\par
Fig. \ref{CumFroll} shows a comparison of the critical rolling friction force of micrometer-sized $\mathrm{SiO_2}$ and $\mathrm{H_2O}$ particles. On average, the ice particles possess a $9.5$ times higher critical rolling friction force than the $\mathrm{SiO_2}$ particles.
\par
From the critical rolling friction force, the specific surface energy of the ice particles can be estimated using the proportionality between $F_{Roll}$ and $\gamma$ (see Eq. \ref{eq0}) and the surface energy of micrometer-sized $\mathrm{SiO_2}$ particles. Measurements of the specific surface energy for micrometer-sized $\mathrm{SiO_2}$ particles yielded values between $\gamma_{SiO_2} = 0.025 \, \mathrm{J \, m^{-2}}$ \citep{Kendall1987} and $\gamma_{SiO_2} = (0.014 \pm 0.002 ) \, \mathrm{J \, m^{-2}}$ \citep{Heim1999}. Thus, a mean value of $\overline{\gamma}_{SiO_2} = 0.020 \, \mathrm{J \, m^{-2}}$ was used for the calculation of the specific surface energy of the micrometer-sized ice particles with Eq. \ref{eq0}. $\xi_{H_2O}$ was then estimated on the assumption that it is equal to $\xi_{SiO_2}$. Using $\xi_{H_2O} = \xi_{SiO_2} = 3.2 \,\mathrm{nm}$, a specific surface energy of $\gamma_{ice} = 0.190 \, \mathrm{J \, m^{-2}}$ was derived for the micrometer-sized ice particles. This result is in very good agreement with the calculated specific surface energy of macroscopic water ice, $\gamma'_{ice} = 0.100 \, \mathrm{J \, m^{-2}} $ \citep{Israel1992}, which was also used in computer simulations of ice aggregate collisions \citep{Wada2007,Wada2008}. In these simulations, the outcome of collisions between different sized dust and ice aggregates, consisting of individual sub-micrometer-sized particles, was investigated.

\section{Conclusion and outlook}\label{Conclusion and outlock}
Two different experimental setups were established to produce ice particles with radii ranging from $a = 0.24\,\mathrm{\mu m}$ to $a = 6.07\,\mathrm{\mu m}$. We found that the mean radii of the produced ice particles are almost identical for both experimental procedures: $a_{0A} = (1.49 \pm 0.79) \, \mathrm{\mu m}$ (setup A) and $a_{0B} = (1.45 \pm 0.65) \, \mathrm{\mu m}$ (setup B). However, preparing smaller ice particles can be conducted by using another nozzle to disperse the water.
\par
However, a comparison of the volume filling factor of the produced ice aggregates composed of micrometer-sized ice particles yields a strong difference between ice aggregates produced with the different setups. Ice aggregates produced with setup A were relatively compact, $\phi_A = 0.72 \pm 0.04$. This quantity is close to the value for hexagonal close packed material, whereas ice aggregates produced with setup B are highly porous, $\phi_B = 0.11 \pm 0.01$, due to the "hit-and-stick" behavior of the slow sedimenting ice particles in the dry, cold gas environment. Due to their high porosity, measurements of the thermophysical properties of the produced ice aggregates, such as the sublimation of porous $\mathrm{H_2O}$ ice and the heat transport connected with the gas diffusion through the porous ice, will be studied in future with a novel constructed experiment \citep{Gundlach2011}, in order to investigate the physical processes connected with the activity of porous icy bodies in the Solar System, such as comets or the surfaces of icy satellites.
\par
The investigation of the critical rolling friction force of micrometer-sized $\mathrm{SiO_2}$ particles showed that the new constructed experiment (setup B) is capable to reproduce the previous published values (see Table \ref{TabComp}). Our measurements of the critical rolling friction force of micrometer-sized ice particles, $F_{Roll,ice}=(114.8 \pm 23.8) \times 10^{-10}\, \mathrm{N}$, revealed an approximately ten times higher critical rolling friction force compared with the $\mathrm{SiO_2}$ particles, $F_{Roll,SiO_2}=(12.1 \pm 3.6) \times 10^{-10}\, \mathrm{N}$. This result implies that the adhesive bonding between micrometer-sized ice particles is stronger than the bonding strength between $\mathrm{SiO_2}$ particles. An estimation of the specific surface energy of micrometer-sized ice particles, derived from the measured critical rolling friction forces, results in $\gamma_{ice} = 0.190 \, \mathrm{J \, m^{-2}}$. A specific surface energy of $\gamma'_{ice} = 0.100  \, \mathrm{J \, m^{-2}}$ \citep{Israel1992} was also used in numerical simulations, in which collision between ice aggregates were studied \citep{Wada2008}. Whether a higher specific surface energy results in a higher sticking threshold, as predicted by \citet{Wada2008}, still needs to be investigated. Hence, collision experiments with micrometer-sized ice particles and different ice aggregates (e. g. different sizes or volume filling factors) will also be conducted to study the sticking threshold of ice particles and the influence of melting or sintering of ice particles during collision in future experiments.

\subsection*{Acknowledgements}
E. Beitz was supported by DFG under grant BL 298/13-1. We thank Sartorius and Millipore for providing us with different filter types.

\bibliographystyle{model2-names}
\bibliography{bib}

\end{document}